\documentclass[preprint,aps,amsmath,amssymb,nofootinbib,showpacs,tightenlines,
showkeys,secnumarabic]{revtex4}
\usepackage{graphicx}
\newcommand{\nn}{\nonumber}

\newcommand{\di}{\displaystyle}

\begin{document}

\title{Synchronous vs. asynchronous dynamics of diffusion-controlled
reactions}

\author{E. Abad}
\thanks{Corresponding author}
\email{eabad@ulb.ac.be}
\affiliation{Center for Nonlinear Phenomena and Complex Systems,
Universit\'e Libre de Bruxelles
C.P. 231, 1050 Bruxelles, Belgium}

\author{Jonathan L. Bentz}
\affiliation{Iowa State University, Department of Chemistry,
Ames, Iowa 50011-3111}

\author{G. Nicolis}
\affiliation{Center for Nonlinear Phenomena and Complex Systems,
Universit\'e Libre de Bruxelles
C.P. 231, 1050 Bruxelles, Belgium}

\author{John J. Kozak}
\affiliation{Iowa State University, Department of Chemistry,
Ames, Iowa 50011-3111}


\begin{abstract}
An analytical method based on the classical ruin problem is
developed to compute the mean reaction time between two walkers
undergoing a generalized random walk on a $1d$ lattice. At each time
step, either both walkers diffuse simultaneously with probability
$p$ (synchronous event) or one of them diffuses while the other
remains immobile with complementary probability (asynchronous
event). Reaction takes place through same site occupation or
position exchange. We study the influence of the degree of
synchronicity $p$ of the walkers and the lattice size $N$ on the
global reaction's efficiency. For odd $N$, the purely synchronous
case ($p=1$) is always the most effective one, while for even $N$,
the encounter time is minimized by a combination of synchronous
and asynchronous events. This new parity effect is fully confirmed
by Monte Carlo simulations on $1d$ lattices as well as for
$2d$ and $3d$ lattices. In contrast, the $1d$ continuum approximation valid
for sufficiently large lattices predicts a monotonic increase of
the efficiency as a function of $p$. The relevance of the
model for several research areas is briefly discussed.
\end{abstract}
\maketitle
\section{Introduction}

Recently, the modelling of diffusion-controlled reactions on
lattices has attracted renewed interest due to synergies with some
research areas like e.g.\ heterogeneous catalysis
\cite{alban,zhdan}, trapping problems \cite{rice}, spin models
\cite{racz}, game theory \cite{fell}, population dynamics
\cite{dick} or biological problems \cite{faum}. An issue common to
these systems is the important role played by the coexistence of
different intrinsic time scales, the lattice characteristics (size
and dimensionality) and many-body effects. The interplay of these
ingredients may strongly affect the efficiency with which
statistical processes such as front propagation on catalytic
substrates, the spread of an infection, kink propagation in
magnetic systems, or exciton trapping in photosynthetic cells take
place. In order to obtain analytical insight for such systems, it
is often necessary to make simplifying assumptions which
nevertheless preserve their main generic features.

In this spirit, a prototypical diffusion-reaction system
consists of two interacting walkers performing nearest neighbor
jumps on a lattice \cite{fish}. The walkers are assumed to react with
each other whenever they meet at the same lattice site or attempt to
exchange positions. Each of such ``encounters'' results in instantaneous
reaction, the process therefore being diffusion-controlled. Regardless of the
particular outcome of the reaction, a measure for its efficiency will
be clearly given by the mean encounter time of both walkers.

In a pioneering work, Montroll investigated a simplified version of the
problem in which one of the walkers is stationary, thereby playing the
role of a fixed trap \cite{mont1}. In a recent article \cite{nickoz1},
 Montroll's results have been extended with the help of a Markovian method to
account for the possibility of simultaneous displacement of the walkers.
In the present work, the problem is further extended to the case in which
the motion of the walkers consists of a random sequence of two different
events: at each tick of the clock, a synchronous event takes place with
probability $p$, i.e., both walkers
hop simultaneously to randomly chosen nearest neighbor sites.
Alternatively, an asynchronous event takes place with probability
$1-p$, i.e. one of the walkers performs a nearest neighbor jump while
the other remains immobile. Thus, the parameter $p$ interpolates
between the one walker plus trap case studied by Montroll ($p=0$) and the case
of two simultaneously moving walkers ($p=1$). A central question we shall
investigate here concerns the influence of the degree of synchronicity $p$
of the walkers and the size $N$ of the lattice in which they are
embedded upon the reaction's efficiency.

Our model may be of interest in several contexts. One can e.g.
easily accomodate it to allow for events in which both
walkers remain immobile in the original reference frame. Thus,
three qualitative different joint events become possible for the
two-walker system, which can e.g. be interpreted as resulting from
a combination of two internal states for each of the walkers,
namely a diffusing and an immobile state. Such two-state random walks
\cite{weiss2,roer,vanden} are frequently used to model
chromatographic \cite{gand} or electrophoretic separation
processes, in which the propagation of charged particles in an
external field may be occasionally stopped by entanglement with
the molecules of the substrate. Such systems, along with hopping
models for transport and recombination of carriers in solids
\cite{nool,fister,schmid}, might provide additional motivation for
considering a generalized version of the random walk.

A different approach suggests to regard the fluctuations in the
diffusivity of the walkers as a result of random fluctuations of
lattice sites switching between a conducting and a stopping state,
whereby the translational invariance of the lattice is
preserved on average. An alternative formulation of the problem in
terms of fluctuating dichotomous barriers between sites also seems
possible \cite{beni}. Such models for dynamic random media are
relevant for the description of several physiochemical processes
like e.g. ligand diffusion in proteins \cite{elber} or proton
migration in water \cite{tucker}.

On the other hand, if one goes back to the original picture of two
dissimilar walkers $A$ and $B$, the model may be regarded as a
schematized starting point for describing the dynamics of exciton
absorption in biological light-harvesting systems \cite{whit}. In
photosynthetic cells, a photon is absorbed by the pigment
molecules (e.g. chorophyll) in the cell and may give rise to an
excited energy state. The exciton hops by resonant energy transfer
through a network or lattice of 200-500 pigment molecules (antenna
system) and can be eventually trapped at a reaction center
\cite{vang}, which is usually considered to be immobile within the
time scale for trapping (a few hundred picoseconds). The exciton
energy is then used to trigger a series of redox processes in the
chain of chemical reactions leading to the production of sugars
and carbohydrates. Thus, one of the time-limiting steps for the
production of oxygen is precisely the absorption of the exciton by
the trapping center. If we allow for a certain mobility of the
reaction center (thereby generalizing Montroll's approach for
exciton trapping), we can identify the latter with a slowly
hopping walker, say the $A$ walker, while the propagating exciton
would play the role of the $B$ walker. As long as the hopping rate
of the $A$ walker remains small, the situation may be identified
with the almost purely asynchronous case (small $p$); this is
normally the case in {\it in vivo} light-harvesting systems.
However, a modification of the physical properties of the antenna
system so as to make exciton propagation slower might, at least in
principle, lead to interesting antiresonance phenomena as observed
in our model.

The paper is organized as follows. In section 2, we define the
two-walker system in detail and show how it can be recast into an
equivalent one-walker system with absorbing sites. In section 3,
we report analytic work on the $1d$ case and identify
the principal differences between the purely asynchronous ($p=0$),
the purely synchronous case ($p=1$) and the mixed case ($0<p<1$). These
results are also supported by Monte Carlo simulations. In Section 4,
similar results are found for the $2d$ and the $3d$ cases by means of
simulations  In section 5, the results are compared with
the predictions of the continuum approximation valid for large
lattices. Finally, section 6 summarizes the main conclusions and
discusses possible extensions.

\section{Formulation of the problem: two-walker vs. one-walker picture}

The starting point is a 1d periodic lattice with $N$ sites and discrete
time dynamics (cf Fig. \ref{felfig1}a). We place two walkers $A$ and $B$
on two distinct lattice sites and let them evolve at each time
step as follows:

\begin{enumerate}

\item with probability $p$, both walkers hop simultaneously to
randomly chosen nearest neighbor sites ({\it synchronous event}).

\item with probability $1-p$, one of the walkers (no matter which one)
remains at rest while the other performs a jump to a nearest neighbor
site ({\it asynchronous event}).

\end{enumerate}

The walkers are assumed to be unbiased, i.e., their jumps are
symmetric. We additionally assume that their jump directions are
completely uncorrelated. An encounter takes place when both
walkers `land' on the same site or attempt to exchange positions.
Each encounter triggers instantaneously an irreversible reactive
interaction, say the annihilation reaction $A+B\rightarrow 0$. The
encounter time can thus be regarded as the characteristic reaction
time governing the diffusion-controlled two-particle annihilation.

In principle, the mean encounter time can be computed for a single initial
configuration of the walkers; in many practical situations, though, one
has little knowledge about the initial conditions. We shall therefore give
preference to a definition of the mean encounter time which contains
an additional coarse graining over all possible initial
configurations. In the sequel, we shall denote this quantity by
$\langle n \rangle$. The smaller $\langle n \rangle$, the more
efficient the reaction will be.

Note that, due to the translational invariance of the lattice, the
physical distinguishability of the walkers is irrelevant
for the computation of $\langle n \rangle$: in other words, it does not
matter which of the walkers $A$ or $B$ hops more often, since
$\langle n \rangle$ depends only on the relative motion of both
walkers, the latter being fully characterized by $p$. Therefore, we
shall assume for simplicity and without loss of generality that $A$ and
$B$ are physically identical walkers and drop the labels $A$ and $B$,
as has been done in Fig. \ref{felfig1}a.

For $p=0$, only reactions through simultaneous site occupancy are possible.
As far as $\langle n \rangle$ is concerned, this case is equivalent to the
one treated by Montroll \cite{mont1}. However, if $p>0$, reaction through
position exchange becomes possible. This technical obstacle makes the
analytical treatment of the problem more difficult.

A first step to circumvent this difficulty is to take advantage of the 
translational invariance of the lattice and reformulate the problem
 in terms of a single walker. To do so, we must switch over
to a new comoving reference frame in which one of the walkers remains
stationary, thereby playing the role of a fixed perfect trap $T$
(Fig. \ref{felfig1}b). Clearly, an asynchronous event in the original
two-walker system is equivalent to
nearest neighbor hopping in the single-walker system, while a
synchronous event results either in a walker's jump by two lattice
sites (if both walkers hop in opposite directions in the original
reference frame) or its remaining at rest (if they hop in the same
direction). In this picture, reaction takes place any time the walker
reaches or overreaches the trap $T$.

We can now take advantage of
the simple lattice geometry and the fact that the trap is perfect
to unfold the $N$-site lattice into an equivalent one with $N+1$ sites
and two perfect traps $T$ sitting at each end site, as shown in Fig.
\ref{felfig2} for $N=7$. This transformation does not of course affect the
characteristics of the walk; a walker jump in anticlockwise
direction will be equivalent to a jump to the left by the same number
of sites in the transformed lattice. Next, we replace each trap $T$ by
two (fictitious) reactive sites $r$ (Fig. \ref{felfig3}). The random walk
will instantaneously terminate when the walker lands on any of
these $r$-sites. We shall therefore occasionally call the $r$-sites
``absorbing'' in what follows.

Let us assign the coordinates $1$ to $N-1$ to each of the
non-absorbing sites, as shown in Fig. \ref{felfig3}. Additionally,
we term the $r$-sites at the left end $-1$ and $0$ and those at
the right end $N$ and $N+1$. An attempt to overcome, say the left trap,
triggered by a synchronous event will be equivalent to a jump from
site $1$ to site $-1$. In contrast, the walker's landing on the
trap can be realized either by an asynchronous event (jump from
site $1$ to site $0$) or by a synchronous event (jump from site
$2$ to site $0$). If the dynamics is purely asynchronous, only one
$r$-site at each end will be needed, since jumps by two sites are
not possible in this case. Thus, events involving position exchange 
can be dealt with by going over to a one-walker picture and 
introducing fictitious absorbing sites. The mean duration of the 
walk averaged over the initial positions $1,\ldots,N-1$ of the 
walker plays the role of $\langle n \rangle$ in the original 
two-walker system. 

A standard approach for the calculation of $\langle n\rangle$ 
is to formulate the problem for the restricted
walk between the absorbing sites in terms of a conditional first
passage problem for an unrestricted walk on an infinite $1d$ lattice
\cite{weiss}. The starting point is the Markovian master equation
\begin{equation}
\label{masteq}
P_{n+1}(j) = \frac{p}{4} \left[ P_{n}(j-2)  +  2 P_{n}(j)  + 
P_{n}(j+2) \right] + \frac{1-p}{2} \left[ P_{n}(j-1)  +  P_{n}(j+1) \right], 
\end{equation}
where $P_n(j)$ is the probability to find the walker at a given site
$j$ in the infinite lattice 
after $n$ time steps. The first term on the right hand side of eq. 
(\ref{masteq}) is
the contribution due to the synchronous events, by which the
walker either remains at rest or it moves (here) two lattice sites
either to the left or to the right. The second term describes jumps 
by one lattice site as a consequence of asynchronous events.  
The mean time to absorption $\langle n \rangle$ can now be viewed
as a first-passage time and computed via a  
generating function approach \cite{weiss}. Specifically, $\langle n \rangle$
can be expressed in terms of the generating function
for the probability of the walker arriving for the first time at a
given $r$-site after a given number of steps without having been
previously in neither of the other three $r$-sites. Unfortunately, 
the resulting expressions for $\langle n \rangle$ are not very transparent 
and their analytical dependence on $N$ and $p$ is not easy to determine. 

Another possibility is to make use of the single step probabilities 
appearing as coefficients in the right hand side of eq. (\ref{masteq}) 
to compute the transition probabilities between the states of the 
underlying absorbing Markov chain. In this approach, $\langle n \rangle$ 
can be related to the row sums over the elements of the
fundamental matrix for the transition to the absorbing state
or, alternatively, to its smallest eigenvalue \cite{kem}. The 
disadvantage of this method is that it requires the inversion
of increasingly large matrices as $N$ becomes large.

Finally, the method we shall further develop here exploits the analogy
of our random walk problem with the classical
ruin problem studied by Feller \cite{fell}. Even though this approach
has the disadvantage of being difficult to generalize to higher
dimensions, it provides an elegant solution for the 1d problem.

\section{Connection with ruin problem and analytical solution}

We first recall briefly the classical gambler's ruin problem. Consider
a single walker (in our setting, this would correspond to the limit
p=0), whose position $z$ is viewed as the capital of a gambler
playing against an adversary whose capital is $N-z$. At each time
step, a trial is made, as a result of which the gambler wins or
loses one euro. Thus, the gambler's winning corresponds to a
nearest neighbor jump of the walker to the right, while losing the
trial corresponds to a jump to the left. The game goes on until
the gambler's capital is reduced to zero or increases to $N$
(absorption of the walker at sites $0$ or $N$). One is
interested in the mean duration of the game $\langle n \rangle_z$ 
when the gambler starts with a given capital $z$. This quantity 
can be shown to be finite as long as 
$N<\infty$ and obeys the following difference equation \cite{fell}:
\begin{equation}
\label{mtdiffeq}
\langle n \rangle_z =\frac{1}{2}\langle n \rangle_{z+1}+
\frac{1}{2}\langle n \rangle_{z-1}+1, \qquad 0<z<N
\end{equation}
with the boundary conditions
\begin{equation}
\label{bcdiffeq}
\langle n \rangle_0=0 \quad \mbox{and} \quad \langle n \rangle_N=0.
\end{equation}
Eq. (\ref{mtdiffeq}) states
that the walker has no memory of where it has been at earlier time steps;
if it is initially at site
$z$, it will either jump to site $z+1$ or to site $z-1$ with
probability $1/2$. Once at any of these sites, the walker will
continue his walk without remembering its previous position $z$.
It is as though the walker started a new walk from $z+1$ or $z-1$
with equal probability, except that the expected value of the mean
time to absorption must be increased by one unit \cite{walsh}.
The boundary conditions (\ref{bcdiffeq}) reflect the fact that, if the
walker is initially placed at a $r$-site, it is immediately absorbed.
In the original two-walker system, this is equivalent to placing both
walkers at the same site.

Eqs. (\ref{mtdiffeq})-(\ref{bcdiffeq}) can be solved by standard
methods, e.g. by writing the general solution as the sum of the
general solution of the corresponding homogeneous equation plus a
particular solution \cite{jord}. One obtains in this way
\begin{equation}
\label{enctspd}
\langle n\rangle_z=z\,(N-z),
\end{equation}
and
\begin{equation}
\label{montenct}
\langle n\rangle =\frac{1}{N}\sum_{z=1}^{N-1} \langle n\rangle_z
=\frac{N(N+1)}{6}.
\end{equation}

The opposite limit of the above ($p=1$), corresponding in our
setting to a purely synchronous motion of {\it two} walkers,
is somewhat less standard. In this case, only jumps by zero or 
two lattice sites may occur. Depending on its initial position, 
the walker may land on
any of the four $r$-sites depicted in Fig. \ref{felfig3}.
Therefore, two additional boundary conditions for the absorbing
sites $-1$ and $N+1$ are needed.

In the gambler's jargon, a jump by two sites means doubling the
stake of each trial, i.e., the player wins or loses two euros with
probability 1/4. Besides, a tie occurs with probability 1/2 (no
jump). The game terminates when either the player or his adversary
reaches or overreaches a capital of $N$ euros. Now, the difference
equation for the duration of the game is a fourth-order one:
\begin{equation} 
\label{syndiffeq}
\langle n \rangle_z =\frac{1}{4}\langle n \rangle_{z+2}+
\frac{1}{2}\langle n \rangle_z+ \frac{1}{4}\langle n
\rangle_{z-2}+1, \qquad 0<z<N.
\end{equation}
The solution of this equation requires four boundary conditions, namely
\begin{equation}\label{synbcdiffeq}
\langle n \rangle_{-1}=0, \quad \langle n
\rangle_0=0, \quad \langle n \rangle_N=0 \quad \mbox{and} \quad
\langle n \rangle_{N+1}=0.
\end{equation}
A particular solution of (\ref{syndiffeq}) is given by
$-\frac{1}{2}z^2$. The characteristic equation of the homogeneous
equation has two double roots $\lambda_{1,2}=\pm 1$. Therefore,
the full inhomogeneous solution reads
\begin{equation}
\langle n\rangle_z=-\frac{1}{2}z^2+c_1+c_2 z+(c_3+c_4 z)(-1)^z,
\end{equation}
Substituting into eq. (\ref{syndiffeq}) one obtains a system of 
equations for $c_1$ to $c_4$ whose solution yields 
\begin{align}
\label{multinz}
\langle n\rangle_z &= \left(z(N-1-z)+N \right)/2   &\mbox{
for $z$ odd, $N$ odd,} \nonumber \\
&=z(N+1-z)/2   &\mbox{ for $z$ even, $N$ odd, } \\
&=\left(z(N-z)+(N+1)\right)/2  &\mbox{ for $z$ odd, $N$ even, } \nonumber \\
&=z(N-z)/2   &\mbox{ for $z$ even, $N$ even.}  \nonumber
\end{align}
The spatially averaged time to absorption $\langle n\rangle$ is 
again easily computed. One obtains 
\begin{align}
\label{evenoddenct}
\langle n\rangle &= N(N+1)(N+2)/12(N-1)  &\qquad \mbox{for $N$ even,}  
\nonumber\\
&=(N+1)(N+3)/12  &\qquad \mbox{for $N$ odd.}
\end{align}

For notational convenience, let us rename the value of 
$\langle n \rangle$ obtained from eqs.
(\ref{evenoddenct}) as $\langle n \rangle^{(1)}$, and the
corresponding result of Montroll for the one-walker problem
(eq. (\ref{montenct})) as $\langle n \rangle^{(0)}$. 
A comparison between $\langle n
\rangle^{(0)}$ and $\langle n \rangle^{(1)}$ is shown in Fig.
\ref{felfig6}. As expected, the synchronous case becomes more
efficient as soon as the lattice gets sufficiently large ($N \ge
5$). In the limit of a very large lattice, it is asymptotically
twice as efficient as the asynchronous case [see Fig.\
\ref{felfig7})].

It is also worth comparing the $z$-distribution of
the encounter time for the case $p=1$ (eq. (\ref{multinz})), with
the earlier known result for $p=0$ (eq. (\ref{enctspd})).
For odd $N$, the spatial profile in $z$ displays some qualitative
similarities in both cases (cf. Figs. \ref{felfig4}a and
\ref{felfig4}b). The highest value of $\langle n\rangle_z$ is
attained at those sites $z$ that maximize the distance $d_z=\mbox{min}
(z, N-z)$ from the closest absorbing site. 
 For $p=0$, the encounter time always
increases with increasing $d_z$, and it becomes maximum when
$z=(N\pm 1)/2$. In contrast, the behavior is no longer strictly
monotonic for $p=1$, since $\langle n \rangle_z$ either increases
{\it or} remains constant as $d_z$ becomes larger, giving rise to
a ``staircase'' profile for high enough values of $N$ (Fig.
\ref{felfig4}b).

For even lattices, there are more marked differences between the cases
$p=0$ and $p=1$. For $p=0$, the discrete spatial profile resembles an
inverted parabola like in the odd lattice case, the maximum now
being located at $z=N/2$. However, if $p=1$, one observes a series of
alternating valleys and peaks in the distribution
(cf Figs. \ref{felfig5}a and \ref{felfig5}b). There are two subcases
here: if $N$ is divisible by $4$, the highest values of $\langle n
\rangle_z$ are attained for $z=(N\pm 2)/2$ (Fig. \ref{felfig5}a).
Otherwise, the maximum value corresponds again to
$z=N/2$ (Fig. \ref{felfig5}b). 

Note that for $p=0$ and even $N$ the time to absorption when
the walker is started at sites $1$ or $N-1$ is smaller than in the purely
synchronous case. An intuitive argument points to the fact that, in the
former case, the walker is absorbed one out of two times after the first
time step, while absorption takes place only one out of four times if $p=1$.
However, this argument should be taken with care, since it fails for odd $N$.
Besides, as we shall see later on, the minimum of 
$\langle n \rangle_1=\langle n \rangle_{N-1}$ corresponds to a 
process with $p>0$.

We finally turn to the general case of a mixed walk. Assume that a given event
is synchronous with probability $0<p<1$ and
asynchronous with probability $1-p$. The difference equation for
$\langle n \rangle_z$ now reads:
\begin{widetext}
\begin{equation} \label{gendiffeq} \langle n \rangle_z =\frac{p}{4}\langle n
\rangle_{z+2} +\frac{1-p}{2}\langle n \rangle_{z+1}
+\frac{p}{2}\langle n \rangle_z +\frac{1-p}{2}\langle n
\rangle_{z-1} +\frac{p}{4}\langle n \rangle_{z-2}+1, \qquad 0<z<N.
\end{equation}
The boundary conditions are again given by
(\ref{synbcdiffeq}). A particular solution of eq.
(\ref{gendiffeq}) is given by $-z^2/(1+p)$. The roots of the
underlying characteristic equation are
\begin{equation} \lambda_{1,2}=1, \quad
\lambda_{3,4}=\frac{-1\pm\sqrt{1-p^2}}{p}.
\end{equation}
The full
solution reads
\begin{equation}
\langle n \rangle_z=-z^2/(1+p)+c_1+c_2
z+c_3\,\lambda_3^z+c_4\,\lambda_3^z,
\end{equation}
where the constants $c_1-c_4$ are now given by the linear
system
\begin{subequations}
\begin{eqnarray}
&& -\frac{1}{1+p}+c_1-c_2+\lambda_3^{-1}c_3+\lambda_4^{-1}c_4 = 0, \\
&&\hspace{1cm} c_1+c_3+c_4 = 0, \\
&& -\frac{N^2}{1+p}+c_1+N c_2+\lambda_3^N c_3+\lambda_4^N \,c_4 = 0, \\
&&-\frac{(N+1)^2}{1+p}+c_1+(N+1)\,c_2+\lambda_3^{N+1}\,c_3+
\lambda_4^{N+1}\,c_4 = 0.
\end{eqnarray}
\end{subequations}
We prefer to omit the rather lengthy analytic form of the coefficients
$c_1$-$c_4$. From the above equations, $\langle n \rangle_z$ and
$\langle n \rangle$ can be explicitly computed. For $\langle n \rangle$
one has an expression of the form
\begin{eqnarray}
\label{enctgen}
 \langle n \rangle&=& -\frac{N(2N-1)}{6(1+p)}+c_1(N,p)+
\frac{N\,c_2(N,p)}{2} \nn \\
&& +\frac{1}{N-1}\left( c_3(N,p)\,[\lambda_3(p)]^{N-1}+
c_4(N,p)\,[\lambda_4(p)]^{N-1}\right).
\end{eqnarray}
\end{widetext}
Table \ref{tabenct} displays the analytical expressions of
$\langle n \rangle$ as a function of $p$ for increasing values of
$N$. These are rational functions of $p$ whose complexity
increases with $N$. For $N\ge 4$ and small $p$, one can obtain a
first-order approximation to the exact solution by
Taylor-expanding eq. (\ref{enctgen}):
\begin{equation} \langle n \rangle
=\langle n \rangle^{(0)}\,\left[\,1+\frac{N-3}{N}\,p+
{\cal O} \left(p^2\right)\,\right].
\end{equation}
Figs. \ref{felfig8}a and \ref{felfig8}b contain several numerical
plots of $\langle n \rangle$, as a function of $p$ for different
values of $N$. For $N=2$, the purely asynchronous case is more
effective than the purely synchronous case \footnote{Obviously, 
the purely asynchronous case has a maximum
efficiency in this case, since the walker is always trapped after
the first step.}. As
synchronicity is turned on, a monotonic increase of $\langle n
\rangle$ as a function of $p$ is observed. For $N=3$, $\langle n
\rangle$ does not depend on $p$. At each individual time step, the
probability of reaction is always $1/2$, regardless of whether
both walkers hop or only one of them. In a sense, the cases $N=2$
and $N=3$ are non-generic, since they only involve a single
symmetry-distinct initial condition. For $N\ge 4$, a new parity
effect appears: if $N$ is even, the most effective process is
observed for an intermediate value of $p$ associated to a minimum
of the function $\langle n \rangle$ in the physically acceptable
$p$-interval $[0,1]$; in contrast, for odd lattices, $\langle n
\rangle$ is a monotonically decreasing function of $p$, i.e., the
most effective process is always the purely synchronous one. For
$N=4$, any intermediate value of $p$ makes hopping more effective
than in both limiting cases, and a minimum of $\langle n \rangle$
is obtained for $p_{min}=2/3$. For higher, even values of $N$, the
minimum is rapidly shifted to the right ($p_{min}\approx 0.86$ for
$N=6$) and the $p$-interval for which processes are more efficient
than the purely synchronous case shrinks dramatically. For large
$N$, $p_{min}$ gets arbitrarily close to $1$ (see Table
\ref{pmintab}).

A series of Monte Carlo simulations for the periodic two-walker system has
been carried out to confirm our analytic results based on the one-walker
description. For two different lattice sizes, namely
$N=7$ and $N=8$, we have performed a series of statistical runs to
compute $\langle n \rangle$, each run thereby comprising a whole set of
symmetric-distinct nonreactive configurations. Due to the large variability
$\langle n^2 \rangle-\langle n \rangle^2$ characteristic of first-passage
problems, a relatively high number of runs was needed ($10^6$) to obtain
an accurate value for $\langle n \rangle$. The $\langle n \rangle$-values
obtained from simulations (with an accuracy of three significant digits)
are listed in Tables \ref{tabsimtheo1} and \ref{tabsimtheo2} for
the cases $N=7$ and $N=8$, respectively. The agreement with the analytical
predictions is good, the maximum observed deviation is off the
theoretical value by about $1\%$ only.

In order to obtain additional insight in the even-odd transition
mechanism, we have studied qualitatively $\langle n \rangle_z$ as
a function of $p$ and $N$ for each single initial position $z$ of
the walker. Let us characterize each site $z$ by its distance
$d_z$ to the closest $r$-site. For not too large
values of $p$, the behavior of the encounter time is roughly the
same for all $z$ values regardless of the parity of $N$, i.e., a
decrease of the encounter time is observed (Figs. \ref{felfig9}a
and \ref{felfig9}b). However, the qualitative $p$-dependence of
$\langle n \rangle_z$ in the large $p$ limit becomes different for
initial positions with even or odd values of $d_z$: for even
values of $N$ and initial positions with odd $d_z$, $\langle n
\rangle_z$ begins to increase sharply, as a result of which a
minimum of the curve is observed (cf Fig. \ref{felfig9}a). Even
though the contribution to the global efficiency $\langle n
\rangle$ arising from the $(N/2)-1$ sites with even $d_z$ decreases
strongly in this regime, this effect is overcome by the increase
of the contribution yielded by the $N/2$ sites with odd $d_z$,
thus giving rise to a net increase in $\langle n \rangle$. In
contrast, the sensitivity to the initial condition is less
systematic and less important for odd values of $N$ (Fig.
\ref{felfig9}b): again, a monotonic decrease is observed for even
$d_z$. Even though minima are still observed for $d_z=1$, for all
other odd values of $d_z$, they are either absent 
\footnote{This is e.g. the case for $N=7$ and $d_z=3$ (not shown
here).} or very flat (see curve for $d_z=3$ in Fig. \ref{felfig9}b).

\section{Monte Carlo results in two and three dimensions}

To complement the analytic $1d$ results for the dependence of
$\langle n \rangle$ on the value of the parameter $p$, we have
also investigated how $\langle n \rangle$ depends on $p$ for
dimensions $2d$ and $3d$ using Monte Carlo simulations. Figures
\ref{2DAsynch} and \ref{3DAsynch} show the dependence of 
$\langle n\rangle$ with respect to $p$. As shown above for $1d$, 
there is distinctively different behavior based on whether $N$ is 
even or odd. In higher dimensions this even-odd effect becomes 
much more pronounced.  It is clear that when $N$ is an even number, 
the maximal efficiency is attained at 
some intermediate value between 0 and 1. Also, one notices that
for the $6\times 6$ square lattice and for all $3d$ even lattices 
which we studied $\langle n \rangle^{(0)} < \langle n \rangle^{(1)}$. 
This is a surprising result which we do not see in the $1d$ case
for lattices with $N>4$. This suggests the existence of a
crossover effect in the large size limit when switching from 
$2d$ to $3d$ lattices, implying that in the former case the
two simultaneously moving walkers are more effective than
two asynchronous walkers, while in $3d$ the opposite holds.
As in the $1d$ case, the value 
of $p_{min}$ in $2d$ and $3d$ tends toward 1 in the limit of large
lattice size, but it is interesting to note that in the largest
$3d$ lattice which we studied ($N=1000$), the difference between
$p=0.999$ and $p=1$ is $\sim 600,$ which is about $30\%$ of
$\langle n\rangle^{(1)}$ in that case. This shows that for even
lattices, a minute amount of asynchronicity allows for a much
greater efficiency.  Similar arguments to those given above for
$1d$ exist for in $2d$ and $3d$ when attempting to determine why such
even-odd behavior arises, and further analysis of this striking
behavior is given in Ref. \cite{bentz}.

\section{Comparison with continuum approximation}

It is instructive to compare the above results in $1d$ with the
continuum approximation valid for large $N$. To do so, consider
the one-walker system with absorbing sites and a fixed lattice
length $L=N \Delta x$, where $\Delta x$ is the intersite distance
(lattice constant). According to this definition, $L$ is the
distance between the inmost $r$-sites. The walker's distance to
site $0$ is $x=z\,\Delta x$. We perform the continuum limit by
letting $\Delta x$ and $\Delta t$ simultaneously go to zero under
the additional requirement that the diffusive combination $(\Delta
x)^2/\Delta t$ tend to a finite constant. Since $L$ is fixed, this
implies that one lets the number of sites $N$ go to infinity while
the spatial and temporal resolutions $\Delta x$ and $\Delta t$ are
scaled as $1/N$ and $1/N^2$, respectively. Let us next replace
$\langle n \rangle_z$ by a function $\langle n\rangle_x$ varying
smoothly in the space interval $[0,L]$. The mean elapsed time
$\langle t \rangle_x$ to absorption will then simply be the mean
number of steps $\langle n\rangle_x$ times the time unit $\Delta
t$. From eq. (\ref{gendiffeq}), we have:
\begin{widetext}
\begin{equation} \label{dynenct} \frac{p}{4}\,\left[\langle t
\rangle_{x+2\Delta x}-2\langle t \rangle_x+ \langle t
\rangle_{x-2\Delta x}\right]+ \frac{1-p}{2}\,\left[\langle t
\rangle_{x+\Delta x}-2\langle t \rangle_x+ \langle t
\rangle_{x-\Delta x}\right]+\Delta t=0
\end{equation}
\end{widetext}
with the boundary conditions 
\footnote{In the case $p=0$, one only has two boundary conditions,
namely $\langle t \rangle_0=\langle t \rangle_L=0$.}
\begin{equation} \label{tbc}
\langle t \rangle_{-\Delta x}=\langle t \rangle_x=\langle t
\rangle_L= \langle t \rangle_{L+\Delta x}=0.
\end{equation}
We now divide eq. (\ref{dynenct}) by $\Delta t$ and expand the
expressions in the brackets in $\Delta x$. Taking the diffusive
limit in the resulting equation yields
\begin{equation}\label{eqDt}
D\,\frac{d^2 \langle t \rangle_x}{dx^2}=-1,
\end{equation}
where the relative diffusion coefficient $D$ is given by
\begin{equation}
D=(1+p)\lim_{\Delta x,\Delta t \to
0}\frac{(\Delta x)^2}{\Delta t} = (1+p)\,D^{(0)}.
\end{equation}
In the rightmost equation, $D^{(0)}$ is the value of the diffusion
coefficient for $p=0$. For $p>0$, $D^{(0)}$ is increased by the
prefactor $1+p$, i.e., the variance of the single-step
probabilities of the random walk. In this limit, the four boundary
conditions (\ref{tbc}) coalesce into two distinct ones, namely
$\langle t \rangle_0=0$ and $\langle t \rangle_L=0$. The solution
of (\ref{eqDt}) which fulfils these boundary conditions is
\begin{equation}
\langle t \rangle_x=\frac{x\,(L-x)}{2 D}.
\end{equation}
The spatially averaged reaction time $\langle t \rangle$ is
obtained by integrating over $x$:
\begin{equation}\label{ratioenct}
\langle t \rangle =\frac{1}{L}\int_0^L \langle t \rangle_x\,dx =
\frac{L^2}{12D}.
\end{equation}
As expected, $\langle t \rangle$ is proportional to the squared
lattice length and inversely proportional to the relative
diffusion coefficient $D$. For the special cases $p=0$ and $p=1$,
this result is recovered by directly taking the diffusive limit in
the discrete solutions (\ref{montenct}) and (\ref{evenoddenct}).
The relation (\ref{ratioenct}) leads to the asymptotic law
\begin{equation} \label{aslaw}
\frac{\langle t \rangle^{(0)}}{\langle t \rangle}=\lim
_{N\to\infty}\frac{\langle n \rangle^{(0)}}{\langle n \rangle}=
1+p,
\end{equation}
where $\langle t \rangle^{(0)}=L^2/(12
D^{(0)})$ and $\langle n \rangle$ is given by eq. (\ref{enctgen}).
Eq. (\ref{aslaw}) shows that in the continuum limit, the
efficiency of the reaction increases linearly with $p$. The
continuum approximation applies for sufficiently large $N$, namely
when the typical displacement of the walker $\Delta l\equiv
\sqrt{1+p}\,\Delta x$ at each time step is small compared to the
lattice length $N\Delta x$. If this condition is not fulfilled,
the approximation gets significantly worse at large values of $p$
(cf Fig. \ref{felfig10}).

A generalization of (\ref{eqDt}) for higher order moments
$\langle t^j \rangle_x$ can be
obtained by writing down the difference equations for the
discrete quantities $\langle n^j \rangle_z$ and taking the diffusive
limit thereof. One then gets the coupled set of equations
\begin{equation}
\label{encthie}
D\,\frac{d^2 \langle t^{j+1}\rangle_x}{dx^2}=-(j+1)\,
\langle t^j\rangle_x,
\end{equation}
Eqs. (\ref{encthie}), not to be further dealt with here, are well known
from the theory of first-passage problems, where they are usually
obtained from the adjoint Fokker-Planck equation for the underlying
diffusion process \cite{weiss,cox}. Again, deviations from the dynamics
dictated by (\ref{encthie}) are expected for small lattices.

\section{Conclusions and outlook}

We have seen that the $1d$ problem of computing the mean reaction
time between two diffusing co-reactants can be reduced to a
trapping problem for a single walker. The latter can be viewed as
a generalized ruin problem, the duration of the game plays thereby
the role of the mean time to absorption.

In the diffusive limit, equivalent to the limit $N\to\infty$ if the
lattice length $L$ is held fixed, the reaction efficiency increases linearly
with $p$, but important deviations are observed for not too large values
of $N$. Beyond the crossover size $N=4$, a new parity effect is observed.
For odd values of $N$, the reaction time still increases monotonically
(but no longer linearly) with $p$, while for even values of $N$,
the efficiency is optimized for an intermediate value $p_{min}<1$.
In higher dimensions, this parity effect is even more pronounced, i.e.
for even lattices there is a drastic increase in efficiency when a tiny 
amount of synchronicity is introduced. In contrast to the $1d$ case, the
effect is enhanced with increasing $N$.  

Let us briefly comment on the $1d$ results from the perspective of
the ruin problem. Assume that one of the gamblers is
successively given an starting capital of $1,2, \ldots, N-1$ euros
at each round, while his adversary gets $N-1,N-2,\ldots, 1$ euros.
Let us further suppose that the gamblers can choose between two
kinds of trials: the stake for the first one is one euro and there is
no tie. In the second trial, either nobody wins or one of the
gamblers wins two euros. According
to our results, for an odd capital $N$, the gamblers minimize the
average playing time if they always make a two-euro bet. However, if $N$
is even, they should make a small amount of one-euro bets in order to
finish the round as soon as possible.

Our work can be extended in many different ways. Perhaps the most
straightforward one is the detailed characterization of the whole
distribution $P(n)$ in terms of $p$ and $N$. A generalization of 
results such as equation (\ref{aslaw}) to higher dimensional integral and
fractal lattices is also of interest, since it may further clarify the role
of dimensionality and the lattice coordination number. According to our
results, in one-dimension two synchronously moving walkers are 
asymptotically twice as efficient as when they hop one
after the other. In a 2d square planar lattice, Kozak {\it et al.} have
shown that the purely synchronous case is $\sqrt{2}$ times more efficient
than the purely asynchronous one in the large
$N$ limit for odd lattices \cite{nickoz1}. The question is
whether or not the relative efficiency of both processes in lattices with
fractal dimension $1<d_f<2$ lies between $2$ and $\sqrt{2}$. Preliminary
calculations on a Sierpinski gasket (with fractal dimension $d_f=1.585$)
seem to indicate an asymptotic relative efficiency higher
than $2$ in this case, despite the fact that the lattice has (up to the three
vertex sites) the same coordination number as a 2d square planar 
lattice \cite{bentz}.
The reason for this may be the important role played by the specific
form of the lattice boundaries, even in the limit of a large lattice.
This may motivate the study of boundary conditions other than periodic
ones for the two-walker system. However, the analytical treatment of
this case is considerably harder, at least in the framework of 
the method of difference equations, since the lattice is no longer 
translationally invariant.

As a further extension of our work, one can also consider more complex
reactive schemes \cite{nickoz2} involving more than two walkers to
study the combined effect of synchronicity and many-body effects.

\begin{acknowledgments}
We are indebted to K. Karamanos, M. Plapp, A. Provata and F. Vikas
for fruitful discussions. The authors gratefully acknowledge
financial support from the Universit\'e Libre de Bruxelles.  This
work is also supported by a NATO Cooperative linkage grant
PST.CLG.977780 and by the European Space Agency under contract
number 90043.  One of us (JLB) acknowledges support from the
Arthur P. Hellwig Award.
\end{acknowledgments}

\newpage
\subsection*{Figure Captions}
\begin{figure*}[ht]
\centering
\includegraphics[width=14cm,height=6cm]{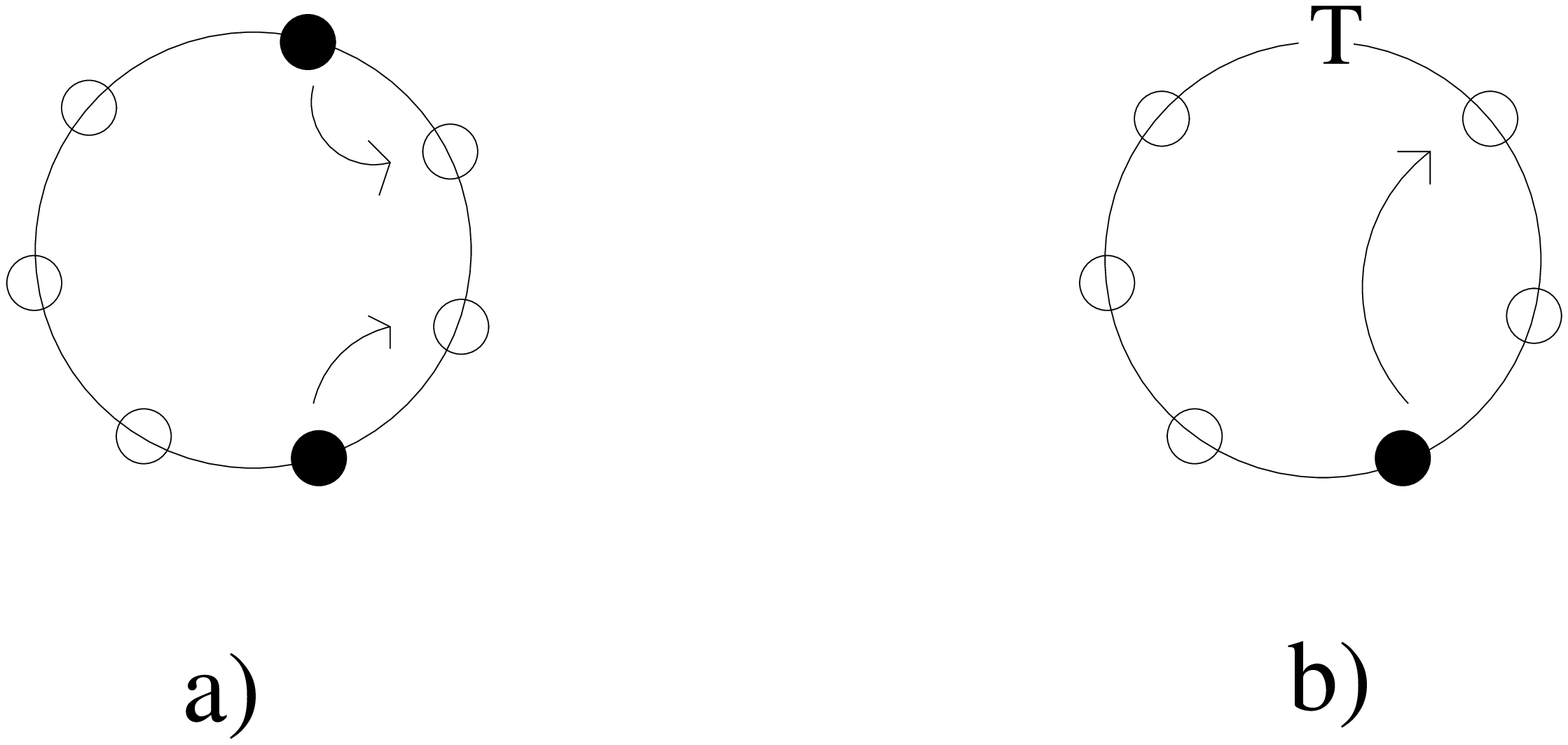}
\caption{\label{felfig1} }
\end{figure*}

\begin{figure*}[ht]
\centering
\includegraphics[width=14cm,height=4cm]{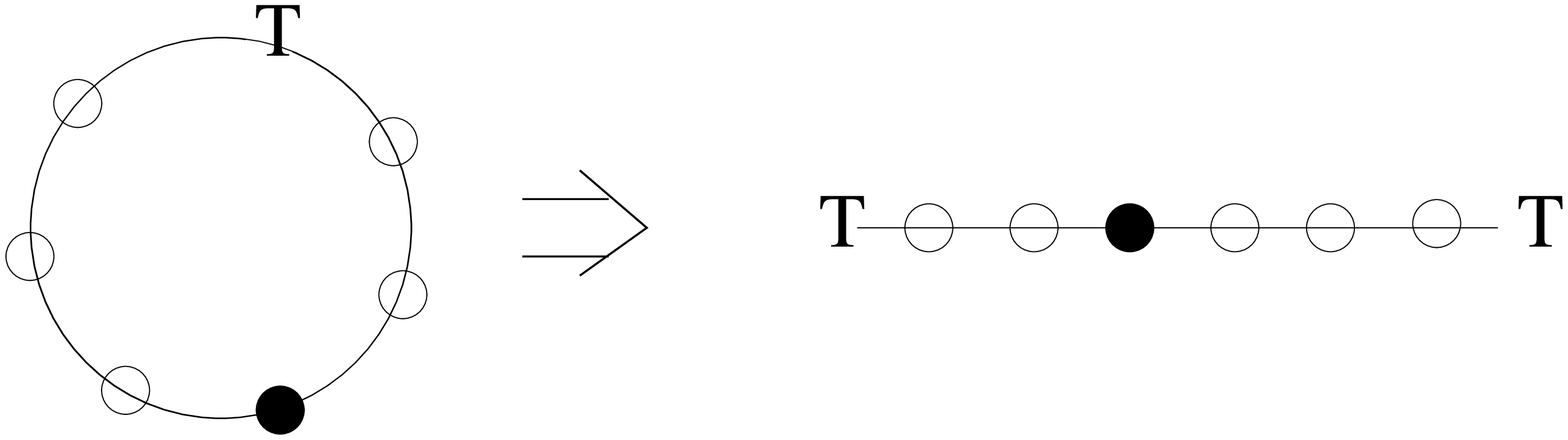}
\caption{\label{felfig2}}
\end{figure*}

\begin{figure*}[ht]
\centering
\includegraphics[width=14cm,height=5cm]{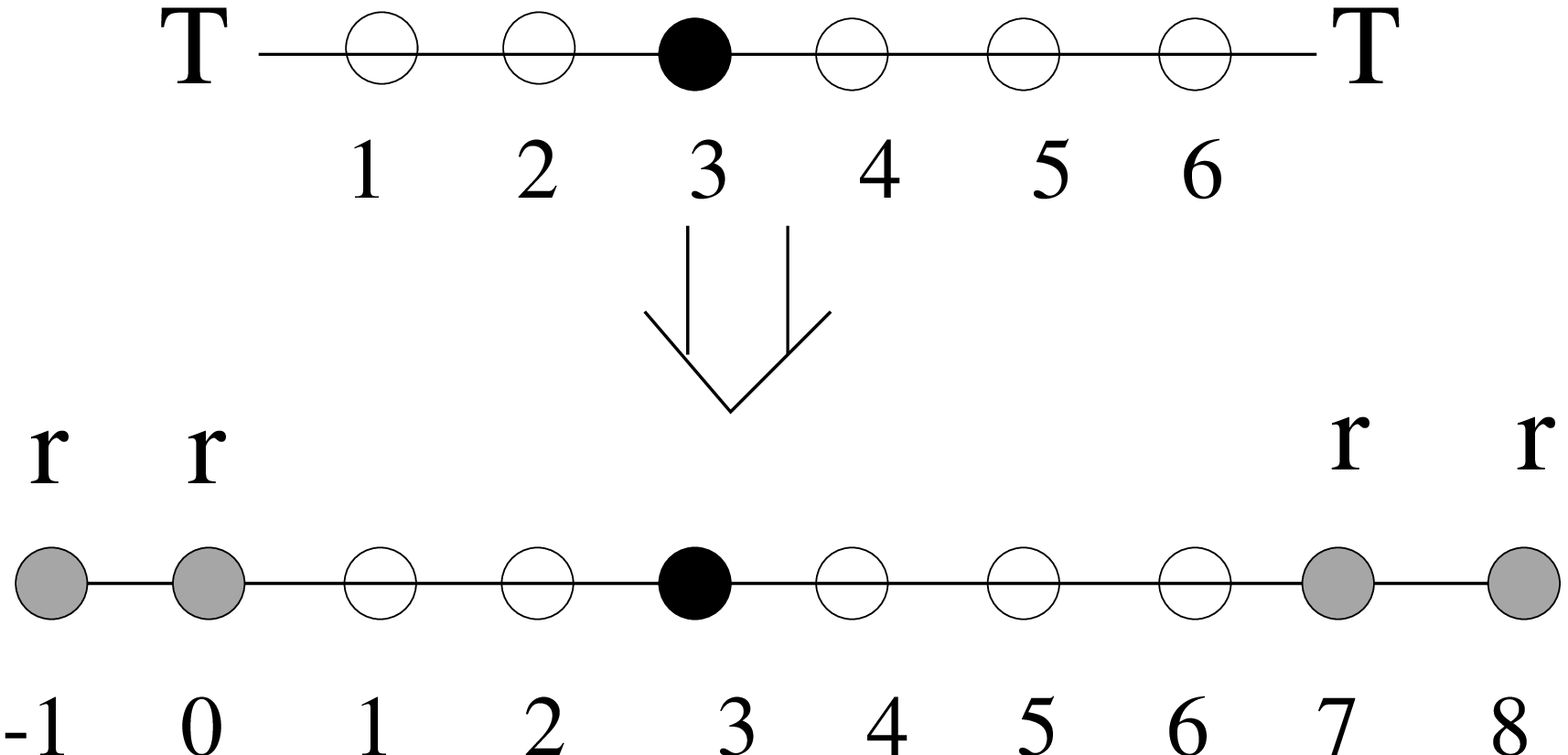}
 \caption{\label{felfig3}}
\end{figure*}

\begin{figure}[ht]
\centering
\includegraphics[width=8cm,height=8cm]{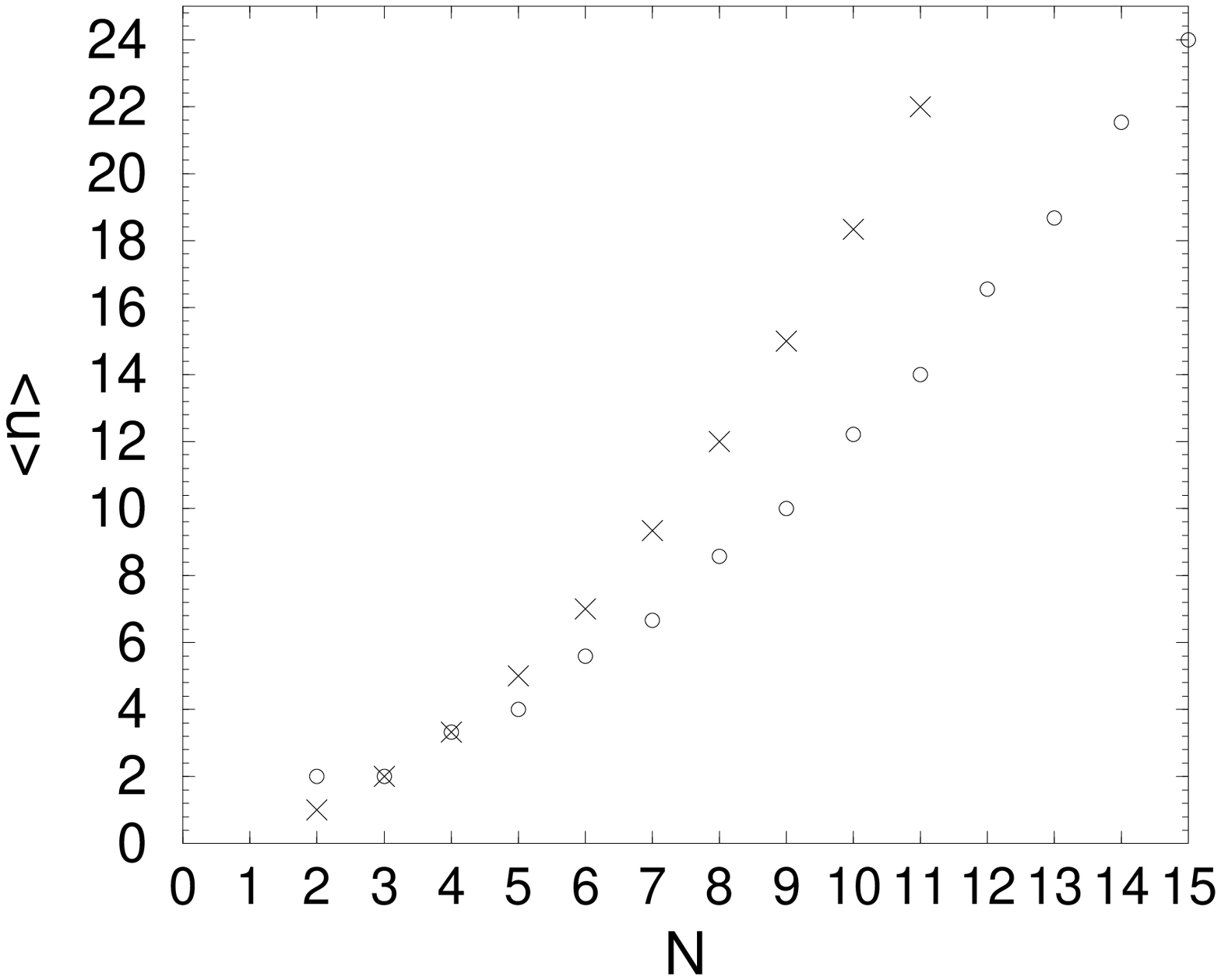}
\caption{\label{felfig6}}
\end{figure}

\begin{figure}[ht]
\centering
\includegraphics[width=8cm,height=8cm]{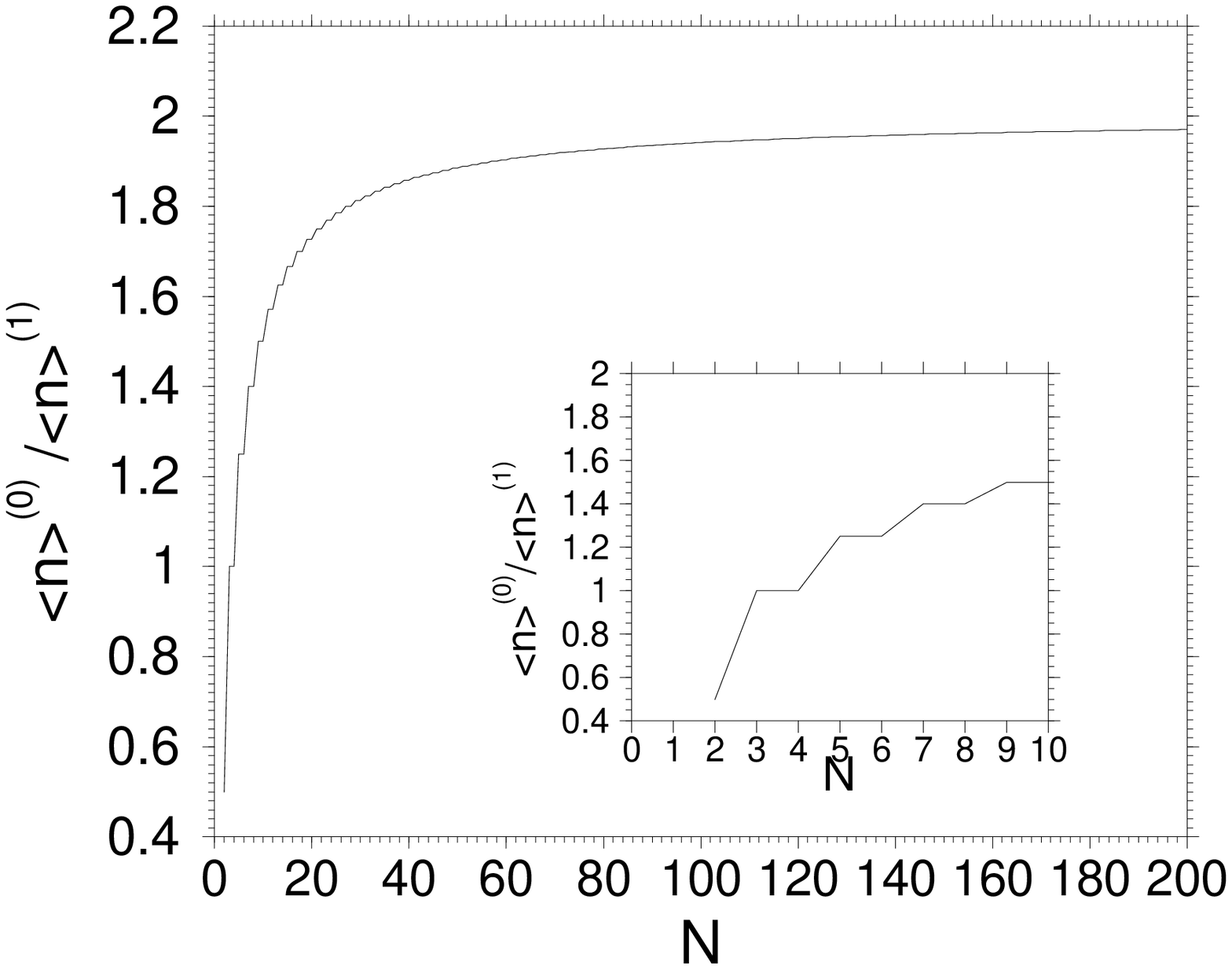}
\caption{\label{felfig7}}
\end{figure}

\begin{figure*}[ht]
\centering
\includegraphics[width=8cm,height=8cm]{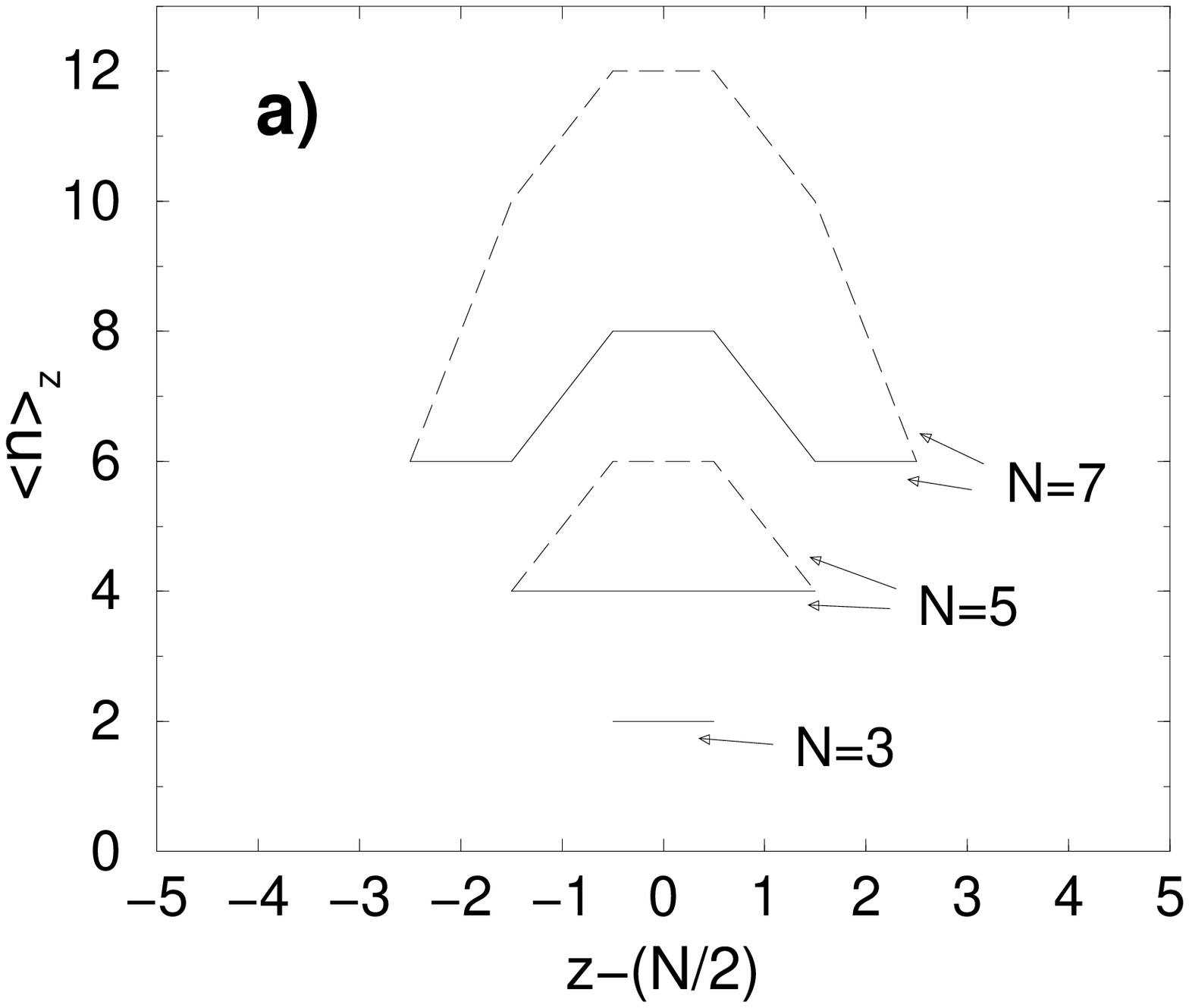}
\includegraphics[width=8cm,height=8cm]{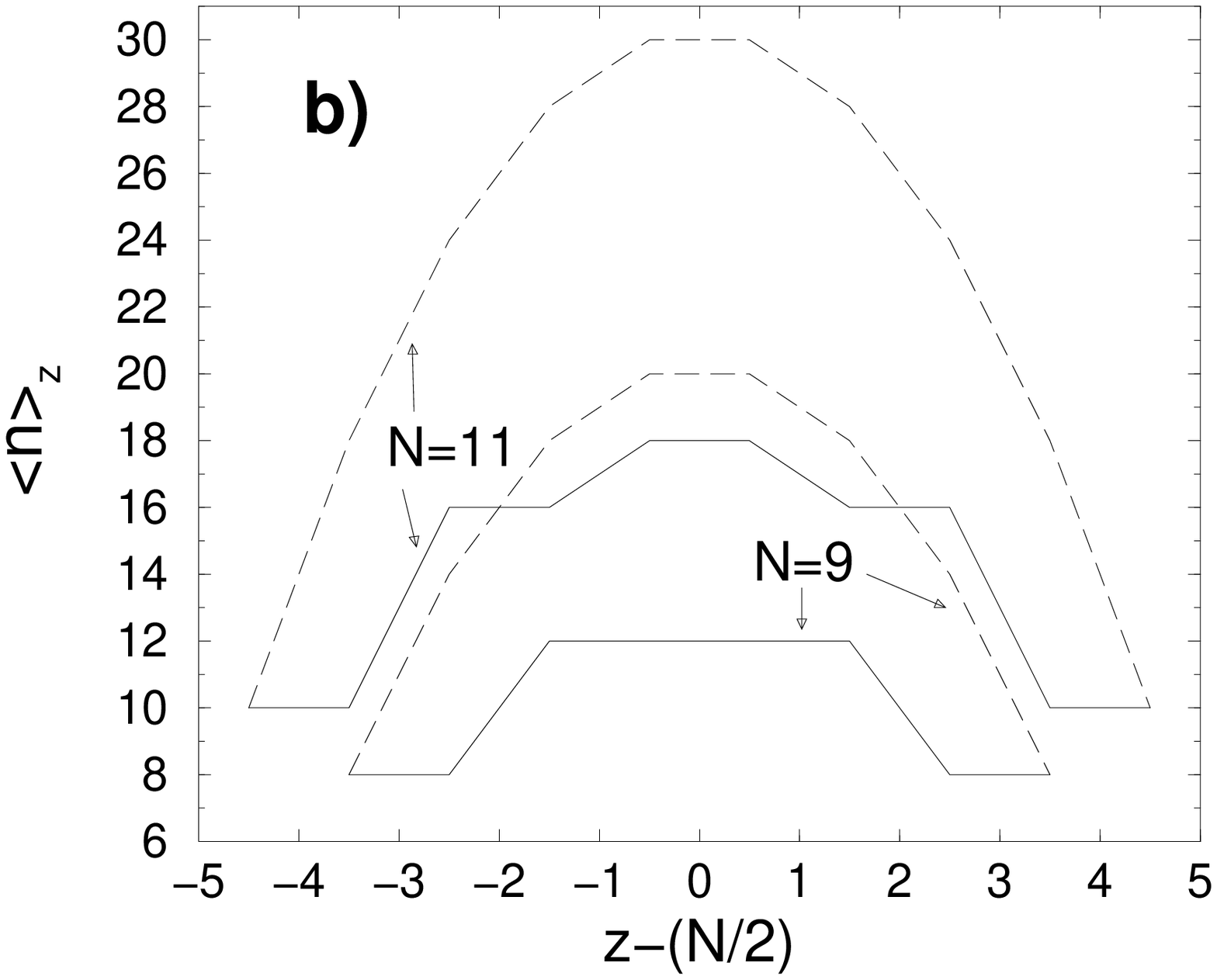}
 \caption{\label{felfig4}}
\end{figure*}

\begin{figure*}[ht]
\centering
\includegraphics[width=8cm,height=8cm]{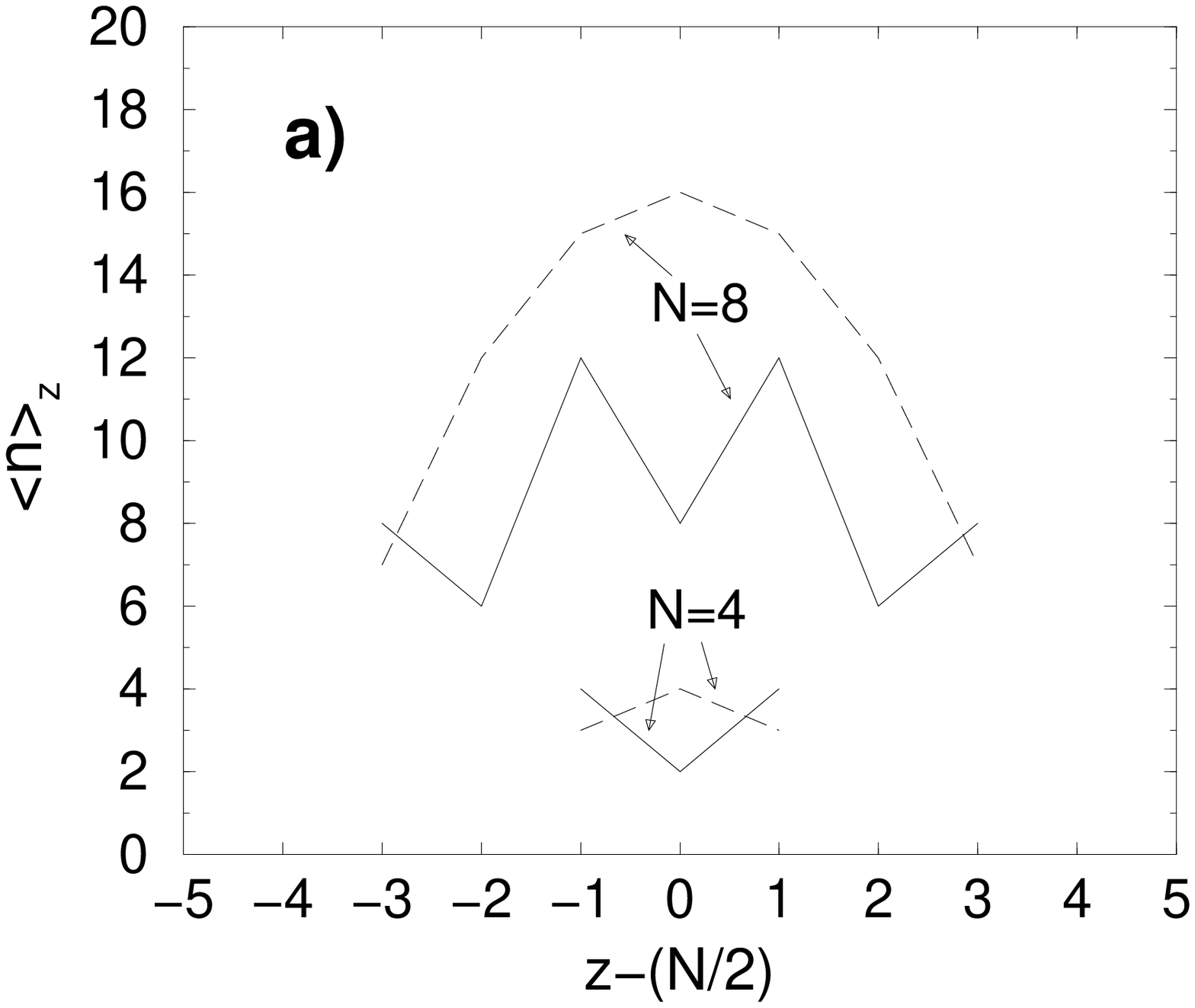}
\includegraphics[width=8cm,height=8cm]{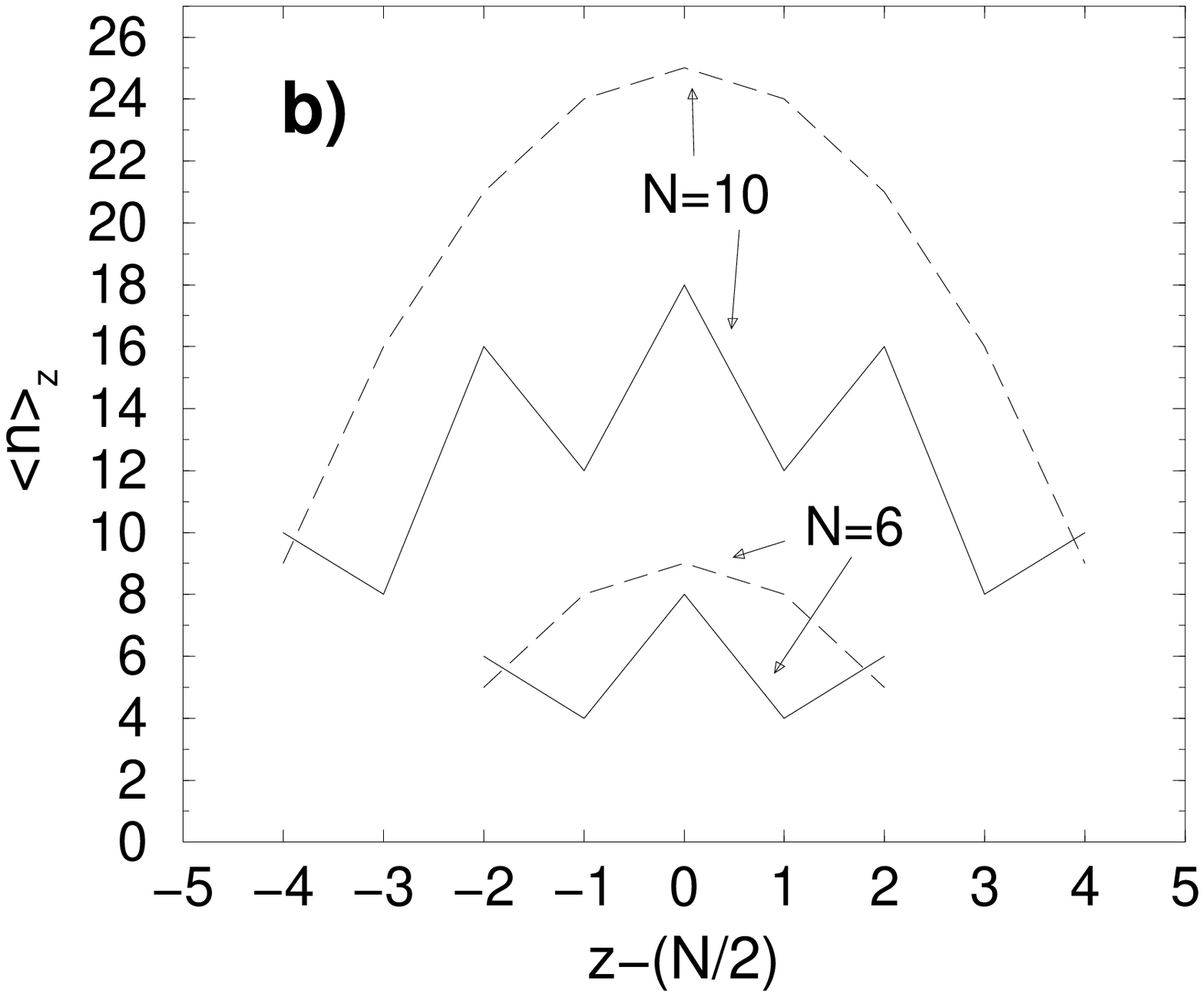}
\caption{\label{felfig5}}
\end{figure*}

\begin{figure*}[ht]
\centering
\includegraphics[width=8cm,height=8cm]{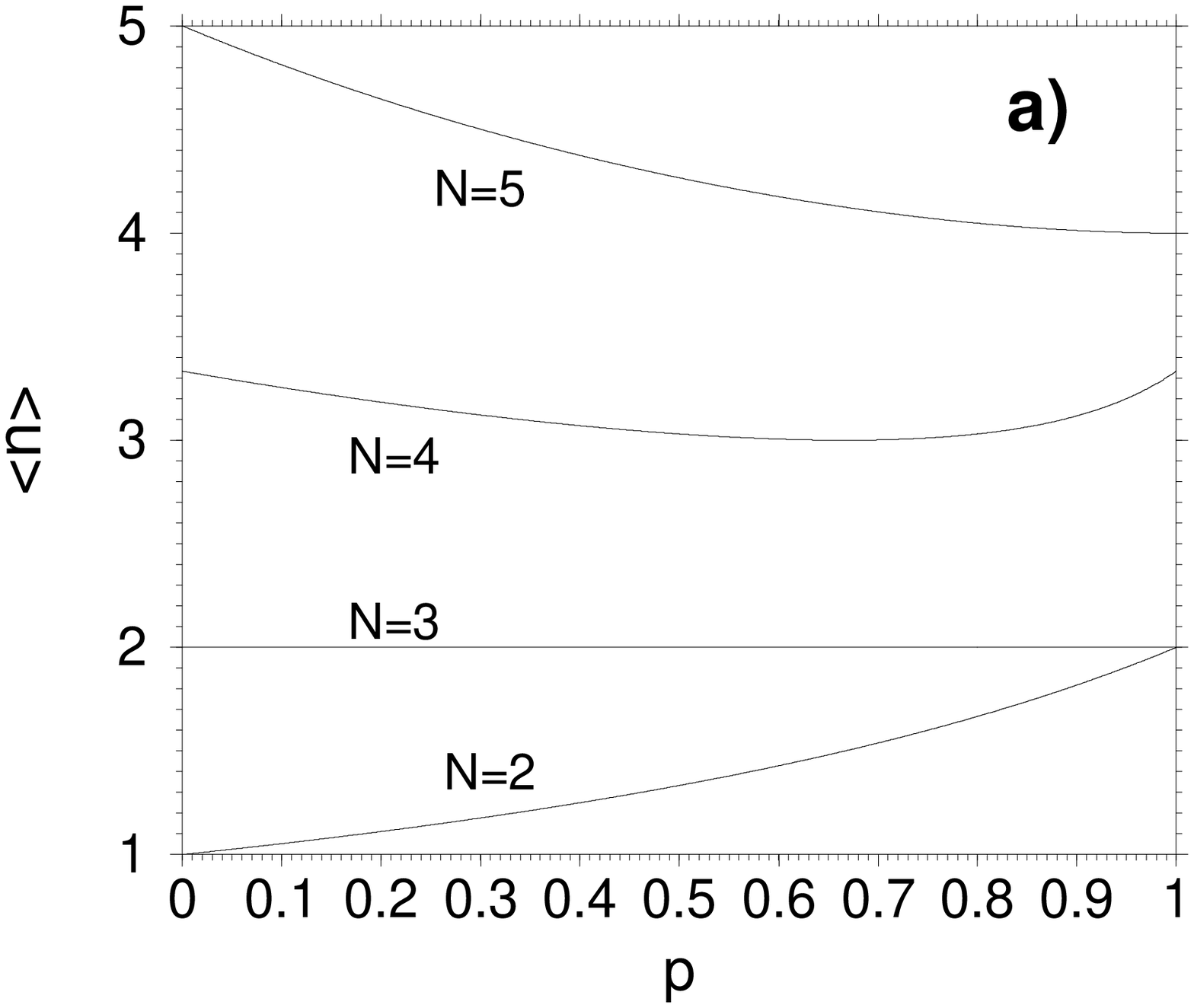}
\includegraphics[width=8cm,height=8cm]{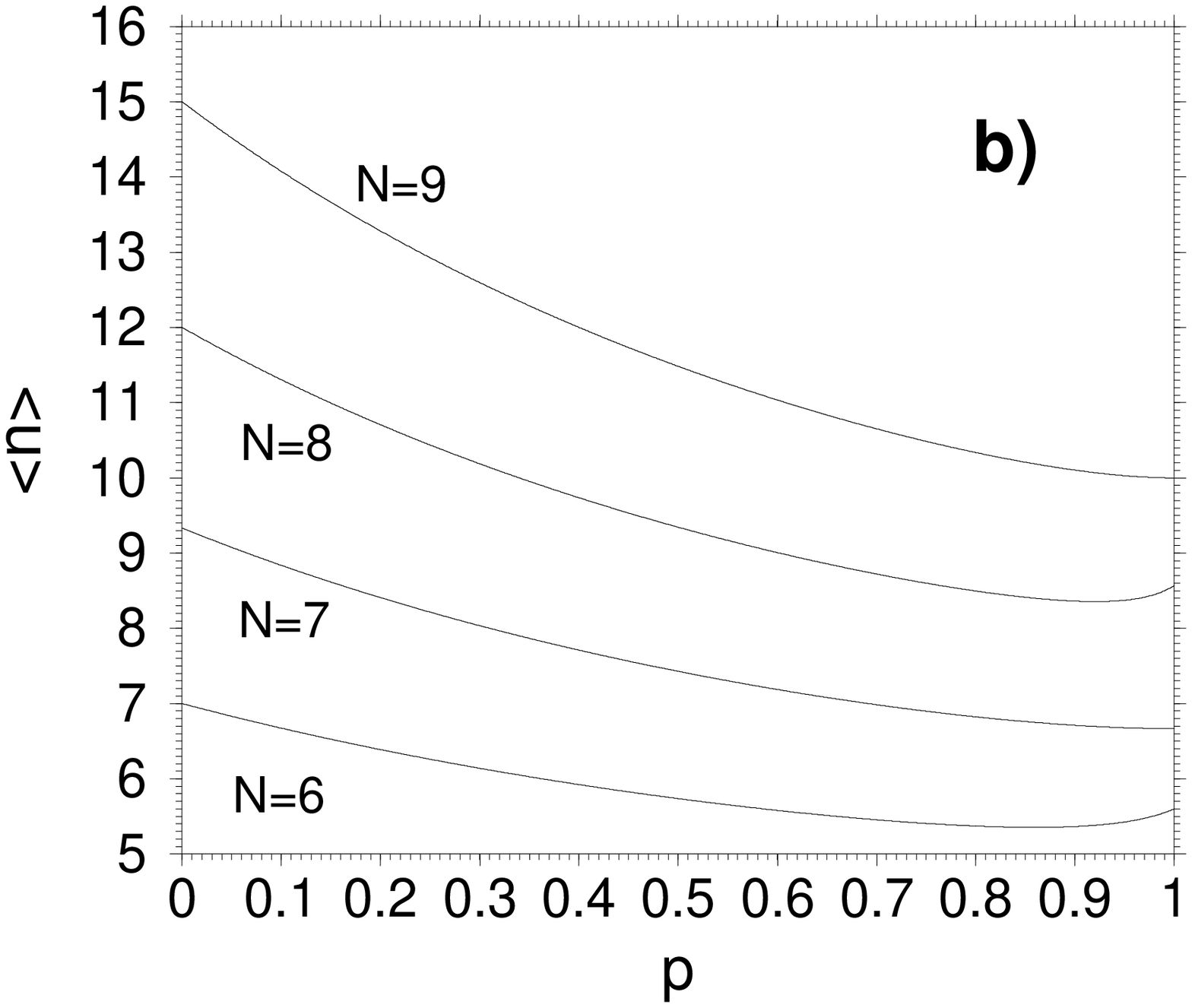}
\caption{\label{felfig8}}
\end{figure*}

\begin{figure*}[ht]
\centering
\includegraphics[width=8cm,height=8cm]{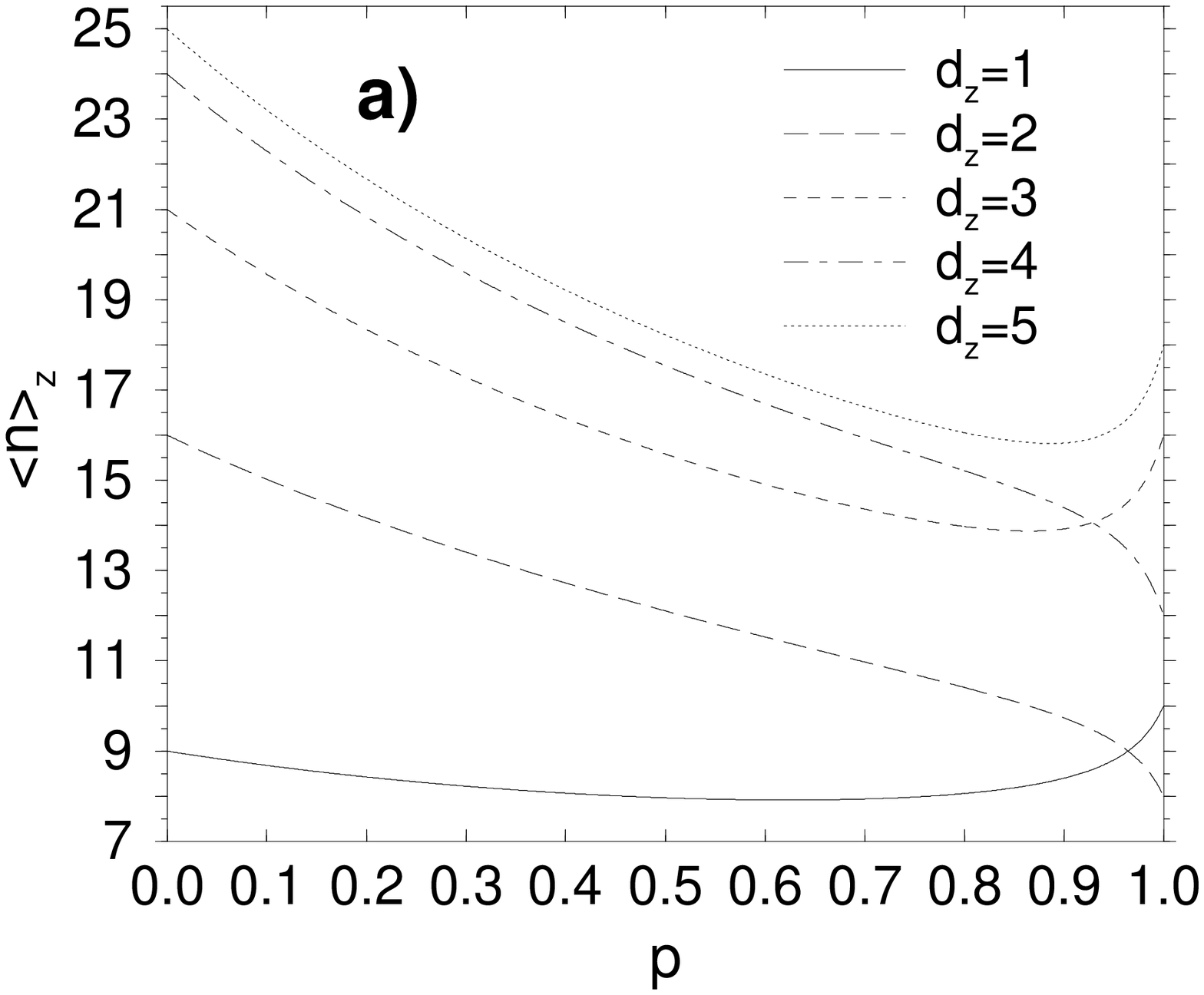}
\includegraphics[width=8cm,height=8cm]{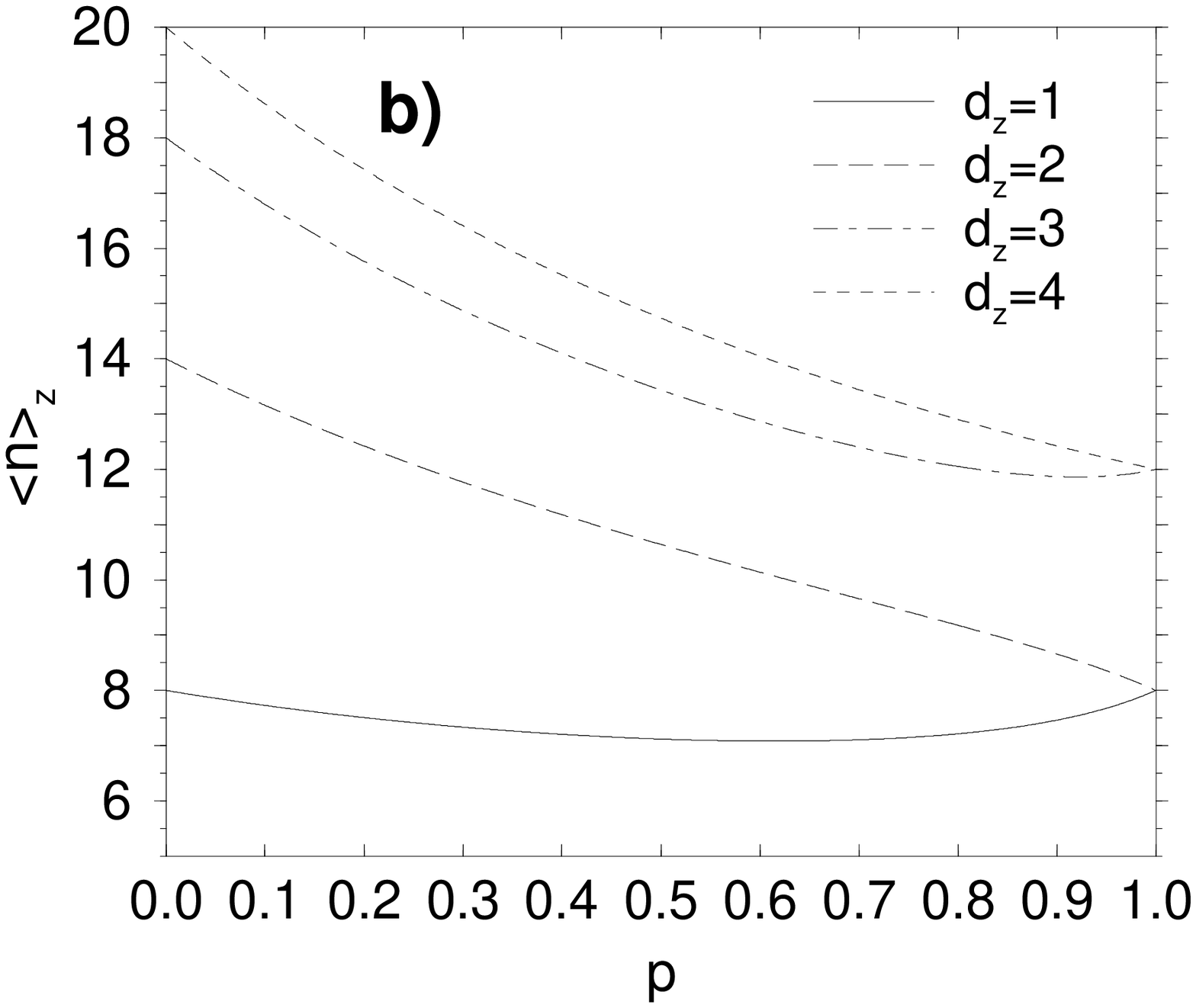}
\caption{\label{felfig9}}
\end{figure*}

\begin{figure}[ht]
    \includegraphics[width=10cm,height=7cm]{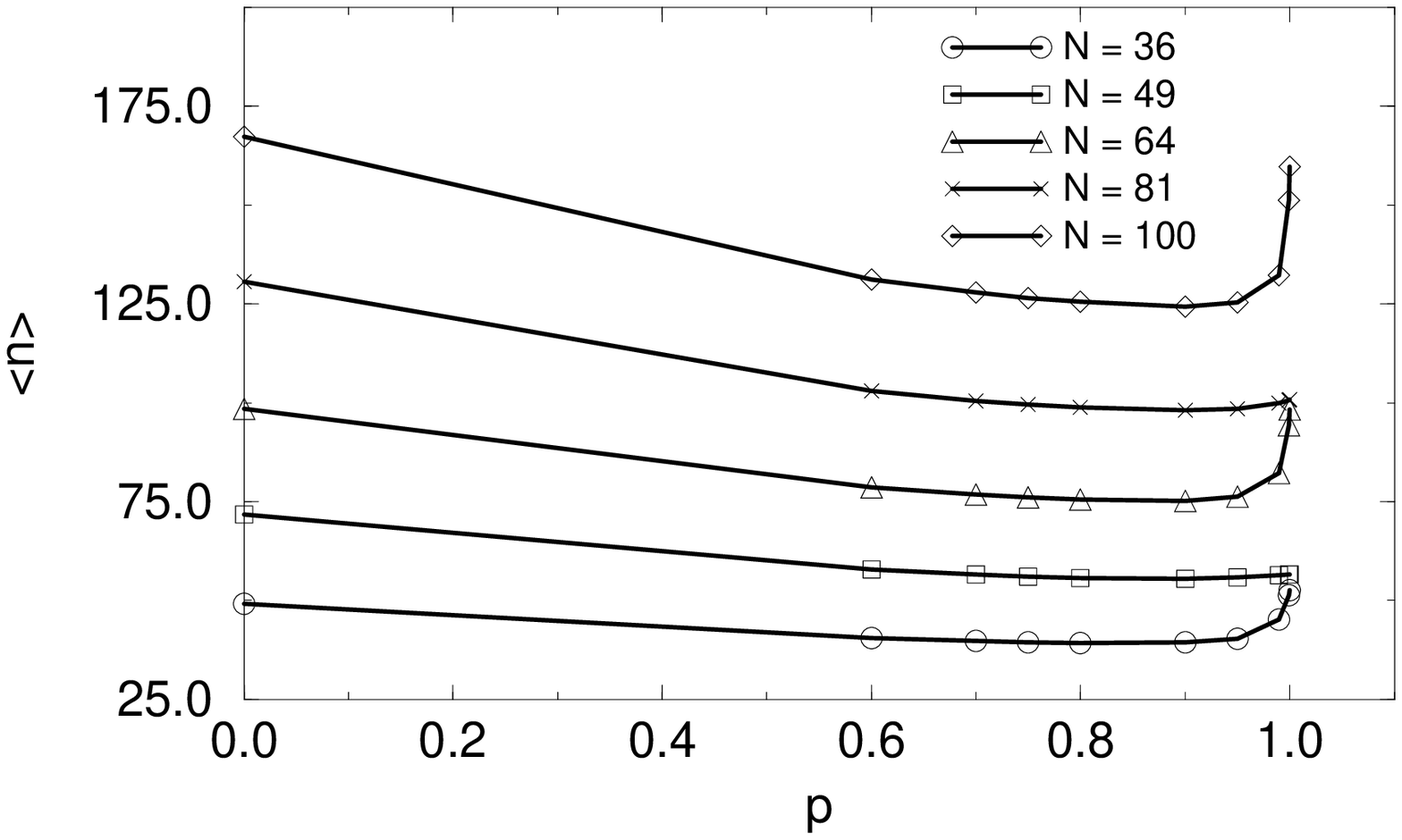}\\
  \caption{}
  \label{2DAsynch}
\end{figure}

\begin{figure}[ht]
    \includegraphics[width=10cm,height=7cm]{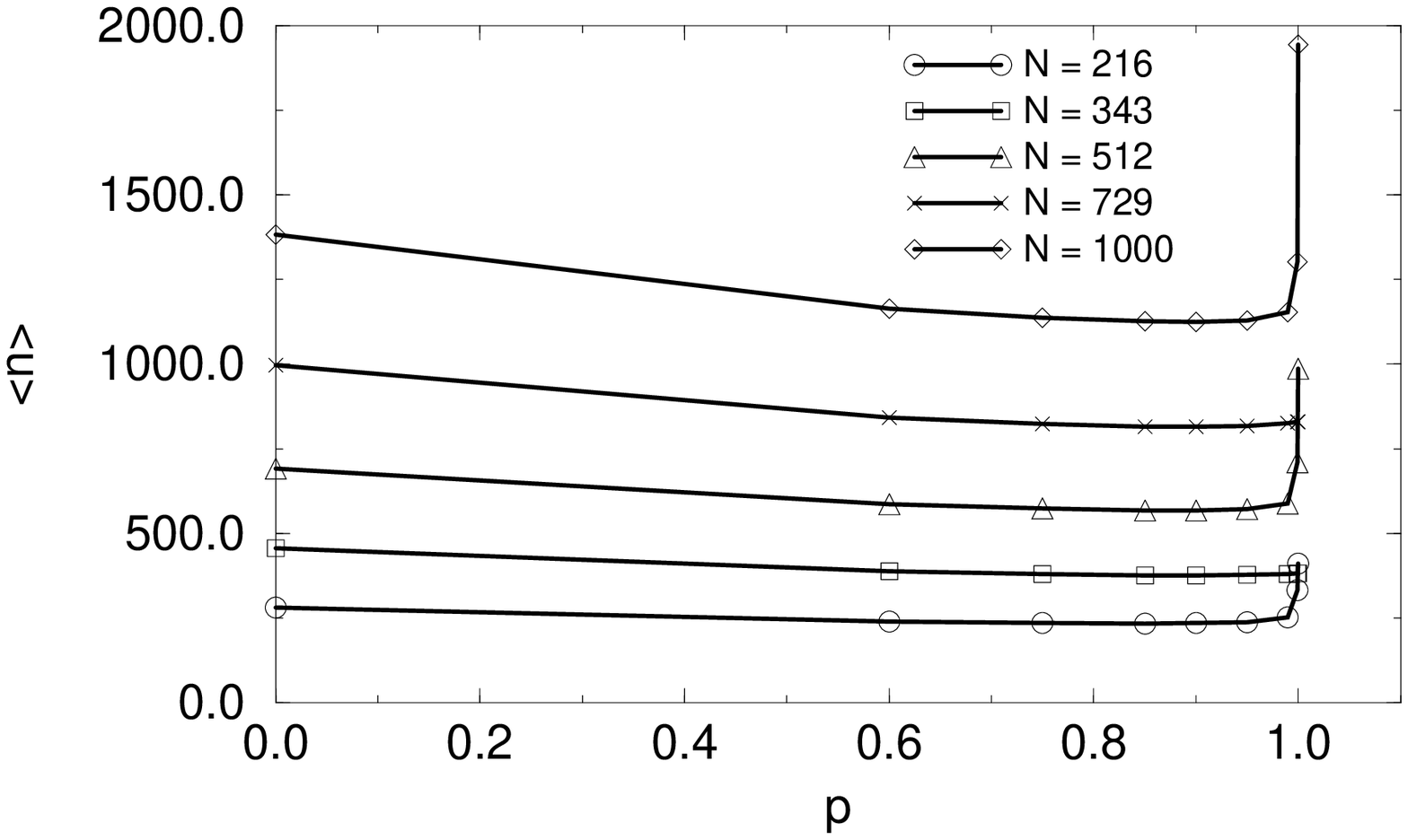}\\
  \caption{}
  \label{3DAsynch}
\end{figure}

\begin{figure}[ht]
\centering
\includegraphics[width=8cm,height=8cm]{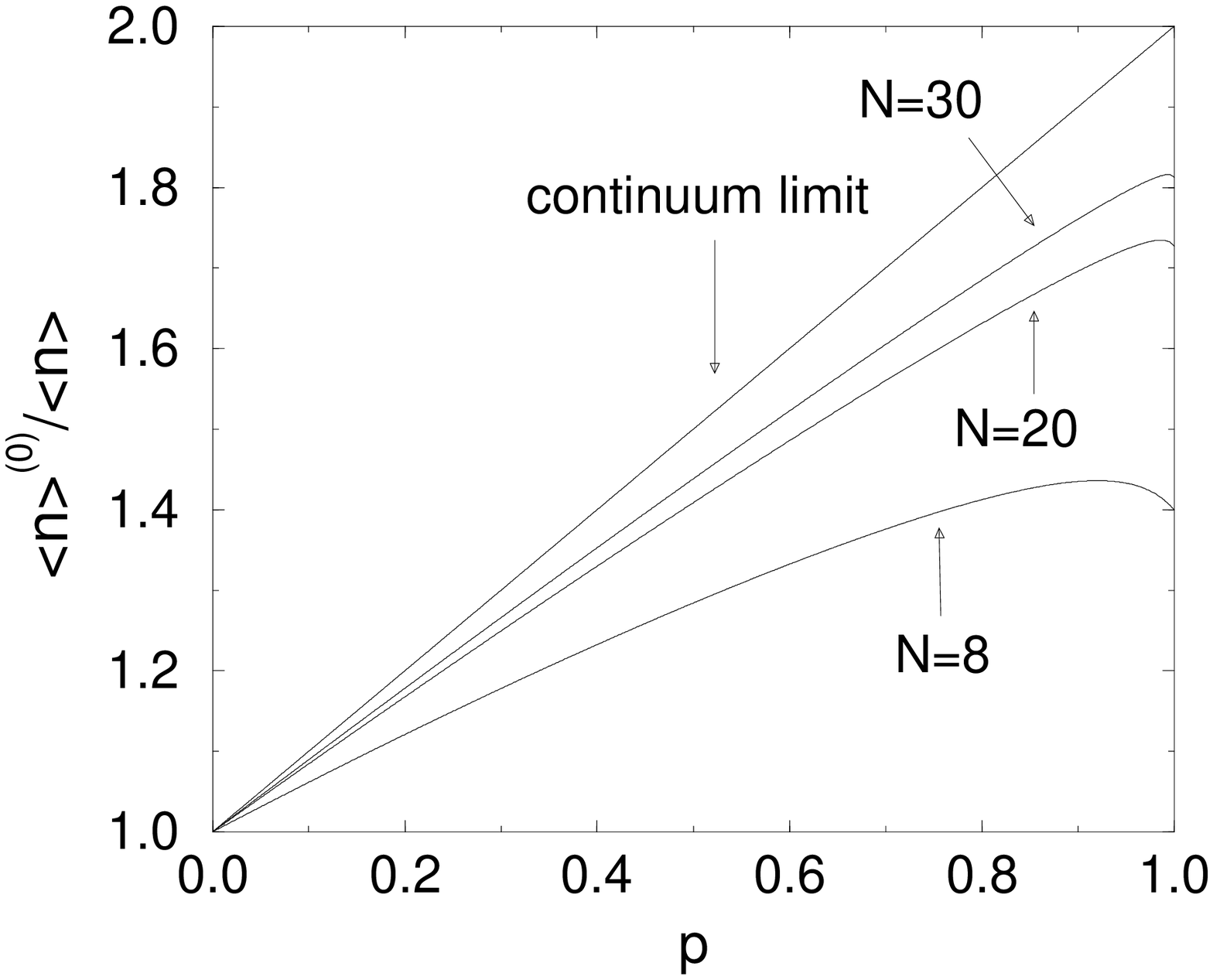}
\caption{\label{felfig10}}
\end{figure}

\noindent Figure 1:  a) two-walker system on a seven-site periodic
lattice (walkers represented by black circles). b) Equivalent
one-walker system with trap. For convenience, the walker labels
$A$ and $B$ in Fig. \ref{felfig1}a have been left out. The arrows
in Fig. \ref{felfig1}a indicate that both walkers perform a
synchronous step. In Fig. 1b this corresponds to a two-site jump
of the walker.

\noindent \\Figure 2:  Lattice transformation for the one-walker
system displayed in Fig. \ref{felfig1}b.

\noindent \\Figure 3:  Replacement of the traps $T$ at each end
site of the transformed lattice by two absorbing sites $r.$

\noindent \\Figure 4: Mean encounter time as a function of the
lattice size for the purely synchronous case $\langle n\rangle=
\langle n\rangle^{(1)}$ (circles) and the purely asynchronous case
$\langle n\rangle= \langle n\rangle^{(0)}$ (crosses).

\noindent \\Figure 5:  Ratio $\langle n \rangle^{(0)}/ \langle n
\rangle^{(1)}$ as a function of $N$. Note that the value of the
ratio is the same for two consecutive odd and even values of $N$.
The inset displays the behaviour for small $N$.

\noindent \\Figure 6:  $z$-distribution of the encounter time in
the cases  $p=0$ (dashed lines) and $p=1$ (continuous lines) for
a) $N=3,5,7$ and b) $N=9,11$. For $N=3$ both cases display the
same flat distribution.

\noindent \\Figure 7:  $z$-distribution of the encounter time for
$p=0$ (dashed lines) and $p=1$ (continuous lines) for a) $N=6,10$
and b) $N=4,8.$

\noindent \\Figure 8:  Mean encounter time as a function of $p$
for a) $N=2,\ldots,5$ and b) $N=6,\ldots,9.$

\noindent \\Figure 9:  a) $p$-dependence of $\langle n \rangle_z$
for all possible values of $d_z$ on a lattice with a) 10 sites b)
$9$ sites.

\noindent \\Figure 10:  Mean encounter time as a function of $p$
for two walkers on a $2d$ square planar lattice with 
periodic boundary conditions.

\noindent \\Figure 11:  Mean encounter time as a function of $p$
for two walkers on a $3d$ cubic lattice with periodic
boundary conditions.

\noindent \\Figure 12:  Ratio $\langle n \rangle^{(0)}/ \langle n
\rangle$ as a function of $p$.

\newpage

\newpage
\begin{table*}[htbp]
\begin{ruledtabular}
\begin{tabular}{cc}
$N$ & $\langle n \rangle $ \\
\hline
2  & $ 2/(2-p)$ \\
3 & 2 \\
4 & $(10/3)\frac{\di 3p-4}{\di p^2+2p-4}$ \\
5 & $4 (2p-5)/( p^2-4)$ \\
6 & $(28/5) (p^2-10p+10)/(p^3-4p^2-4p+8) $ \\
7 & $(4/3) (p^2+8p-14)/(p^2-2)$ \\
8 & $(12/7) (13p^3+6p^2-126p+112)/((p-2)(p^3+6p^2-8))$ \\
9 & $10 (2p^3-5p^2-16p+24)/((p^2+2p-4)(p^2-2p-4))$ \\
10 & $(22/9) (7p^4-76p^3+16p^2+288p-240)/(p^5-6p^4-12p^3+32p^2
+16p-32)$ \\
\end{tabular}
\end{ruledtabular}
\caption{\label{tabenct} Analytic expressions for $\langle n
\rangle$.}
\end{table*}

\begin{table}[htbp]
\begin{ruledtabular}
\begin{tabular}{c}
$N$\hspace{3.5cm}$p_{min}$ \\
\hline
 2 \hspace{3cm} 0.0000 \\
 4  \hspace{3cm} 0.6667 \\
 6  \hspace{3cm} 0.8596 \\
 8  \hspace{3cm} 0.9204 \\
 10 \hspace{3cm}0.9483 \\
 12 \hspace{3cm}0.9636 \\
 14 \hspace{3cm}0.9729 \\
 16 \hspace{3cm}0.9791 \\
\end{tabular}
\end{ruledtabular}
\caption{\label{pmintab} Values of $p_{min}$ with 4-digit
accuracy.}
\end{table}

\begin{table}[htbp]
\begin{ruledtabular}
\begin{tabular}{ccc}
$p$ & $\langle n \rangle^{sim}$ & $\langle n \rangle^{theo}$ \\
\hline
 0.2 &8.410 &8.408163\\
 0.5 &7.430 &7.428571\\
 0.8 &6.826 &6.823529\\
 1 &6.667 &6.666667\\
\end{tabular}
\end{ruledtabular} \caption{\label{tabsimtheo1} Theoretical vs.
simulation value of $\langle n\rangle$ for $N=7$.}
\end{table}

\begin{table}[htbp]
\begin{ruledtabular}
\begin{tabular}{ccc}
$p$ & $\langle n \rangle^{sim}$ & $\langle n \rangle^{theo}$ \\
\hline
 0.2 &10.709 &10.706177\\
 0.5 &9.345 &9.344538\\
 0.8 &8.495 &8.496241\\
 1 &8.572 &8.571429\\
\end{tabular}
\end{ruledtabular}
\caption{\label{tabsimtheo2}  Theoretical vs. simulation value of
$\langle n\rangle$ for $N=8$.}
\end{table}

\end{document}